# Spectral localization of single-nanoparticle plasmons through photonic substrate engineering

Shihao Feng,[1,2,6] Juan-Feng Zhu,[3,6] Xinyi Fan[1,6], Jindong Ai,[1] Wenjie Zhou,[3] Guangxin Liu,[3] Jing-Feng Liu,[5] Renming Liu,[1,2*] Wei Li,[4,*] Lijun Guo[1], and Lin Wu[3,*]

[1]*Henan Key Laboratory of High Efficiency Energy Conversion Science and Technology, Henan International Joint Laboratory of New Energy Materials and Devices, School of Physics and Electronics, Henan University, Kaifeng 475004, China.*

[2]*Institute of Quantum Materials and Physics, Henan Academy of Sciences, Zhengzhou 450046, China.*

[3]*Department of Science, Mathematics and Technology, Singapore University of Technology and Design, 8 Somapah Road, Singapore 487372, Republic of Singapore.*

[4]*State Key Laboratory of Optoelectronic Materials and Technologies, School of Physics, Sun Yat-sen University, Guangzhou 510275, China.*

[5]*College of Electronic Engineering and College of Artificial Intelligence, South China Agricultural University, Guangzhou 510642, China.*

[6]*These authors contributed equally to this work.*

**Corresponding to:** liurm@henu.edu.cn; liwei373@mail.sysu.edu.cn; lin_wu@sutd.edu.sg

Surface plasmon resonances (SPRs) are crucial for confining light beyond the diffraction limit, yet heavy metal losses often limit their spectral localization. Here, we propose a practical strategy for enabling the spectral localization of single-nanoparticle SPRs through photonic substrate engineering, which creates distinct optical pathways (OPs) to tailor the electromagnetic environments around plasmonic nanoparticles. By analyzing the multiplication factor spectrum of the projected local density of states, we can trace and control these OPs, enabling strong spatial and spectral confinement of single-nanoparticle SPRs. Simulations reveal that a photonic crystal substrate can reduce the mode volume by fivefold and boost the quality factor by over 80 times compared to a metal nanoparticle on a dielectric substrate. Proof-of-concept experiments using two types of leaking Fabry–Pérot photonic substrates demonstrate active manipulation of SPRs in both "open" and "closed" OP states. This multidimensional photonic substrate engineering establishes a customizable platform for single-nanoparticle plasmonics, potentially transforming applications that were previously limited by spectral localization.

**Introduction**

Surface plasmon resonances (SPRs) are highly regarded for their exceptional ability to confine light below the diffraction limit[1,2], attracting interest in fields such as information technology[3,4], sensing[5,6], nonlinear optics[7,8], super-resolution imaging[9,10] and biomedicine[11,12]. However, ultrafast decay of excited plasmons is accompanied by heavy losses, posing challenges to "spectral localization" of SPRs. This limitation severely compromises the performance of many plasmonic metal nanodevices[13,14], notwithstanding their advantages of ultrafast optical response, substantial field enhancement, and ease of fabrication.

Spectrum shaping, accomplished through constructive and destructive interference of SPRs, is a widely used strategy for enhancing spectral localization, typically resulting in narrower linewidths. However, significant damping of SPRs often persists[15]. Band structure engineering provides an effective approach for attaining spectral localization with low dissipation, commonly realized in periodic structures such as photonic crystal cavities[16,17], Moiré photonic crystals[18,19], and bound states in the continuum[20-23]. The emergence of plasmonic metasurfaces[24-27] has further enabled effective spectral localization, characterized by resonant modes with high quality factors ($Q$-factors). While these methods are effective, they typically depend on **large-area ordered structures**.

A key challenge is achieving such high spectral localization for SPRs in **single** metal nanostructures due to heavy losses. One viable approach involves hybridizing plasmonic nanostructures with optical microcavities, creating "optoplasmonic" systems[28-44]. Examples include plasmonic nanoparticles paired with Fabry-Pérot (FP) microcavities[29,30] and nanoantennas coupled with whispering-gallery[31-33] or photonic crystal microstructures[34-37]. These configurations utilize high-$Q$ photonic modes to enhance spectral localization, but implementing them presents great challenges. Precise alignment of nanoantennas within photonic structures is critical, necessitating advanced techniques[34,35,37]. Additionally, existing models often overlook the influence of the surrounding vacuum reservoir, resulting in a **fragmented understanding** of phenomena observed in these optoplasmonic systems, such as spectral localization [29,32,39-43] and Fano-resonance destruction[30,31,33,35-38] depending on resonance conditions.

In this work, we present a practical strategy that leverages photonic substrate engineering to achieve high spatial and spectral localization of single-nanoparticle SPRs, eliminating the need for precise positioning within photonic structures. Our **universal theoretical framework** elucidates the coupling between plasmons, photonic modes, and the vacuum reservoir (Fig. 1a), offering a **unified understanding** of the phenomena previously observed in optoplasmonic systems. The carefully designed photonic substrate establishes unique optical pathways (OPs, radiative channels) to the vacuum reservoir, creating a tailored electromagnetic (EM) environment. When a small metal nanoparticle (MNP) is situated in this EM environment (Fig. 1b), its SPRs couple to the vacuum reservoir both directly and indirectly through high-$Q$ photonic modes. By controlling the OPs ("open" or "closed"), we can effectively manipulate the SPRs, which shows great promise for developing high-performance on-chip plasmonic nanodevices, including efficient single-nanoparticle lasing[43], single-photon emission[45,46], ultrasensitive biosensing[47,48], and entangled photon generation[49].

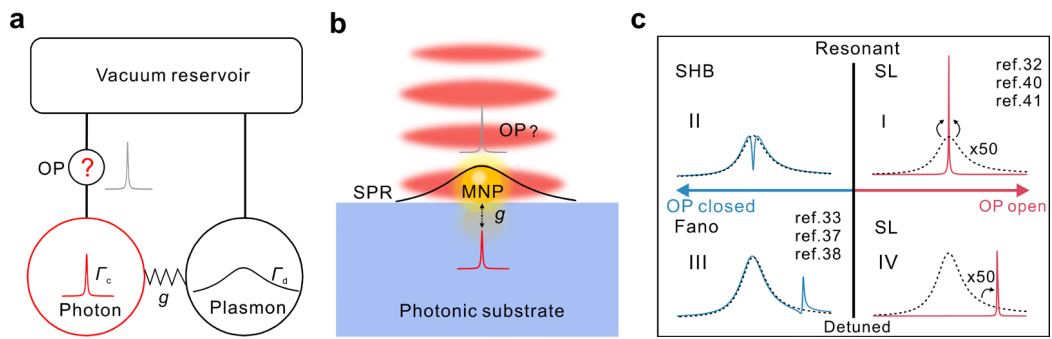

**Fig. 1 Spectral localization (SL) of single-nanoparticle plasmons via photonic substrate**

**engineering. a** Framework illustrating the coupling between a lossy plasmonic mode, a high-Q photonic mode, and the vacuum reservoir through a controlled OP. **b** Schematic of plasmonic SL governed by the OP created by the photonic substrate. **c** Absorption spectrum of the SPR mode for a single MNP on a conventional dielectric substrate (dashed black) versus 4 categories of engineered photonic substrates, reflecting "closed" (solid blue) or "open" (solid red) OP, and resonance conditions between the plasmonic and photonic modes: resonant (upper) or detuned (lower).

## Results

### Universal theoretical framework

We develop a theoretical framework that clarifies the coupling between plasmons, photonic modes, and the vacuum reservoir (Fig. 1a). We assume the small MNP supports a well-defined bosonic dipolar plasmonic mode, described by bosonic annihilation ($d$) and creation ($d^+$) operators, while neglecting higher-multipole modes. For this system, driven by an external weak excitation field ($E_0$), the perturbation Hamiltonian of the plasmonic mode can be written as[50]:

$$H_{dri} = \boldsymbol{\mu}_d \cdot (\boldsymbol{E}_0 + \sum_\alpha \boldsymbol{E}_\alpha) \cdot (d^+ + d) \quad (1)$$

where $\boldsymbol{\mu}_d$ is the transition dipole moment of the SPR mode, $\sum_\alpha \boldsymbol{E}_\alpha$ represents the total electric fields (EFs) from the OPs. Typically, the intensity of the weak excitation field is much weaker than that of OPs ($E_0 \ll E_\alpha$), allowing us to ignore $\boldsymbol{E_0}$ in Eq. (1). According to Fermi's Golden rule, the absorption spectrum of the SPR mode can be expressed as:

$$\sigma(\omega) \propto -\text{Im} \frac{A_d \langle i|dd^+|i\rangle}{\varepsilon_{fi}-\omega-i0^+} = -\text{Im}\left\{A_d \cdot \langle\langle d; d^+\rangle\rangle_{\omega+i0^+}\right\}, \quad (2)$$

where $A_d = |\boldsymbol{\mu}_d \cdot \sum_\alpha \boldsymbol{E}_\alpha|^2 \propto F_m(\boldsymbol{r}, \omega, \hat{\boldsymbol{\mu}}_d) \cdot \rho_0$, and $F_m(\boldsymbol{r}, \omega, \hat{\boldsymbol{\mu}}_d) = \rho(\boldsymbol{r}, \omega, \hat{\boldsymbol{\mu}}_d)/\rho_0$ is the multiplication factor (M-factor) of the projected local density of states (PLDOS), $\rho(\boldsymbol{r}, \omega, \hat{\boldsymbol{\mu}}_d)$, along the dipole orientation $\hat{\boldsymbol{\mu}}_d$[51]. Here, $\rho_0$ represents the PLDOS in vacuum. The two-time correlation function $\langle\langle d; d^+\rangle\rangle_{\omega+i0^+}$ can be derived using the retarded Zubarev Green function formalism[52] and the Heisenberg equation of motion. Finally, the absorption spectrum of the MNP embedded in this engineered electromagnetic environment can be obtained (see Supplementary Note 1 for details):

$$\sigma(\omega) \propto -F_m(\boldsymbol{r},\omega,\widehat{\boldsymbol{\mu}}_d).\operatorname{Im}\left\{\hbar\omega-\varepsilon_d+i\frac{\Gamma_d}{2}-\sum_{j=1}^{N}\frac{g_j^2}{\hbar\omega-\varepsilon_{cj}+i\frac{\Gamma_{cj}}{2}}\right\}^{-1}, \quad (3)$$

where $g_j$ is the coupling strength between the SPR mode and the j$^{\text{th}}$ photonic mode, $\varepsilon_d$ ($\varepsilon_{cj}$) and $\Gamma_d$ ($\Gamma_{cj}$) denote the energy and decay rate (*i.e.*, the damping linewidth) of the SPR mode and the j$^{\text{th}}$ photonic mode, respectively. We note that the scattering spectrum of the SPR mode can also be described by Eq. (3) when probing the optical response from the same plasmonic channel (*i.e.*, the metal nanoparticle), assuming that differences in intensity are negligible[53]. As shown in Fig. 1a, Eq. (3) reveals that the spectral characteristics of the SPR mode, particularly its damping linewidth, can be effectively controlled through the engineered OPs. This is quantitatively captured by the M-factor spectrum $F_m(\boldsymbol{r},\omega,\widehat{\boldsymbol{\mu}}_d)$ evaluated at position $\boldsymbol{r}$, which serves as a direct indicator of the local electromagnetic environment. The M-factor can be accurately obtained through numerical simulations (see Methods and Supplementary Note 2 for details).

Consequently, OPs generated through photonic substrate engineering (Fig. 1b) offer an efficient approach for achieving spectral localization of the SPR modes, resulting in sharper resonances (solid red lines) compared to conventional dielectric substrates (dashed black line), for both resonant (quadrant I) and detuned (quadrant IV) scenarios (Fig. 1c). In the absence of OPs, the term $F_m(\boldsymbol{r},\omega,\widehat{\boldsymbol{\mu}}_d)$ in Eq. (3) can be treated as a constant or effectively neglected. This simplification leads to phenomena such as spectral hole-burning (SHB) (quadrant II in Fig. 1c, resonant) or Fano resonance destruction (quadrant III in Fig. 1c, detuned), as reflected in the observed SPR modes of the MNPs (solid blue lines).

These "open" or "closed" OP scenarios provide us a framework for understanding the diverse phenomena of SPRs reported in optoplasmonic systems[28-44]. For instance, with a closed OP, when the SPR mode is detuned from cavity modes, Fano resonance destruction phenomenon may occur[30,31,33,35-38], while significant SHB of the plasmonic mode can be achieved when they are in resonance (quadrant II in Fig. 1c), similar to that seen in laser-induced bleaching[55,56] and quantum plasmonic systems[57,58].

Interestingly, even with a large detuning, spectral localization can still be achieved if there is an open OP around the MNP (quadrant IV in Fig. 1c), akin to the resonant case (quadrant I in Fig. 1c)[29,32,39-43]. This suggests that the OPs created by photonic substrate engineering are key to unlocking the spectral localization of single-nanoparticle plasmons and provide deep insights into the hybridization in optoplasmonic systems. Next, we present a photonic substrate design that theoretically demonstrates SPR with an ultranarrow linewidth of below 1 nm. We also showcase proof-of-concept experiments for both open and closed OPs to illustrate all cases in the 4 quadrants.

**Designing photonic substrates for ultrasharp (<1 nm) SPR**

We first present a photonic crystal substrate design that numerically demonstrates single-nanoparticle SPRs with ultrahigh spectral localization and an exceptionally narrow linewidth below 1 nm. The photonic crystal substrate consists of a broadband Fabry-Pérot (FP) microcavity and a photonic crystal-guided resonator (PCGR) (Fig. 2a). The structural parameters of this FP-PCGR photonic substrate are $P$ = 400 nm, $P_{f1}$ = 140 nm, $P_{f2}$ = 162 nm, $t_{Au}$ = 100 nm, $t_{SiO_2}$ = 400 nm, $t_{TiO_2}$ = 148 nm, and $d$ = 40 nm. The FP microcavity is defined between the bottom surface of the TiO$_2$ layer and the top surface of the underlying Au film, with an effective cavity length equal to $t_{SiO_2}$ = 400 nm. This configuration generates a unique OP in the EM environment above the photonic crystal substrate (Fig. 2b), characterized by a Lorentzian peak in the M-factor $F_m$ spectrum, centered at 697.5 nm, with a maximum intensity and an ultranarrow damping linewidth of approximately 0.55 nm (solid red, Fig. 2c; Supplementary Figs. S1-S3). This peak corresponds to a similar dip observed in the reflection spectrum (solid black) of the bare FP-PCGR photonic substrate.

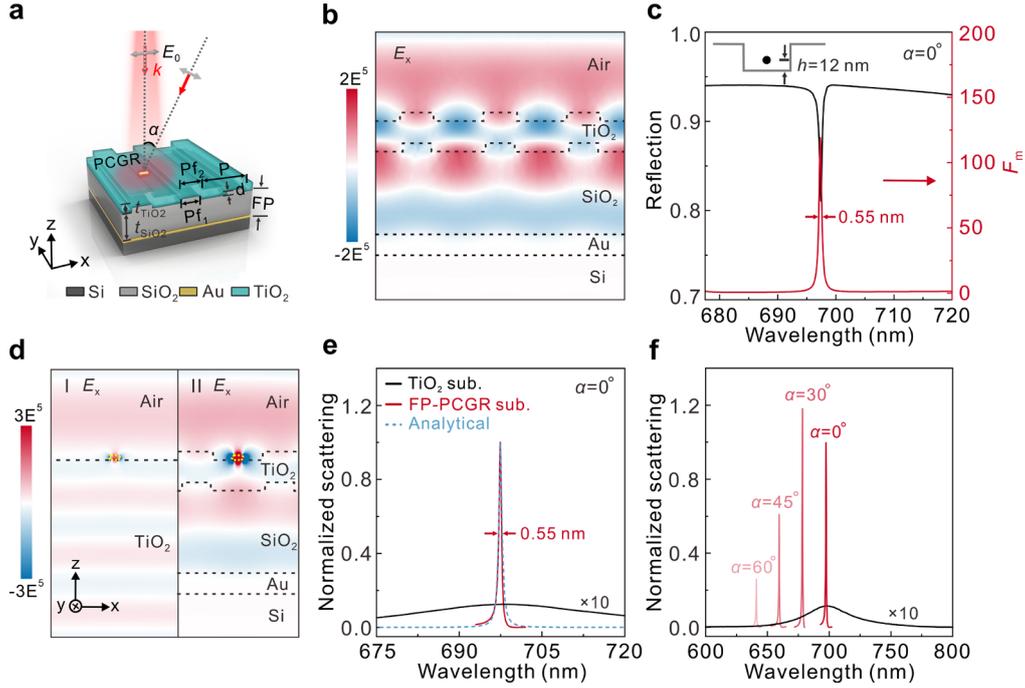

**Fig. 2 Simulated spectral localization for AuNR on the FP-PCGR photonic crystal substrate.** **a** Schematic of an AuNR placed on the groove of a photonic crystal substrate comprising a FP cavity and a photonic crystal-guided resonator, illuminated by TM-polarized light at an incidence angle α. **b** The cross-sectional (x-z) view of the EF distribution ($E_x$) for the bare FP-PCGR substrate without the AuNR. **c** Simulated reflection spectrum of the bare FP-PCGR substrate (solid black) and the M-factor $F_m$ spectrum (solid red) at $h$ = 12 nm above the groove surface (the designated position for the AuNR). **d** The cross-sectional (x-z) view of $E_x$ around the AuNR (54 nm in length and 12 nm in radius) on (I) a bulk dielectric (TiO₂) substrate and (II) the FP-PCGR substrate. **e** Normalized scattering spectra of the AuNR simulated on the FP-PCGR substrate (solid red) and the TiO₂ substrate (solid black), with the dashed blue line showing the analytical result from Eq. (3). **f** Normalized scattering spectra of the AuNR on the FP-PCGR substrate simulated at incidence angles α = 0°, 30°, 45°, and 60°. The black curve shows the spectrum on a TiO₂ substrate. All spectra are normalized to the α = 0° case. Material dispersions are provided in Supplementary Fig. S3.

We then position a gold nanorod (AuNR) on the **top surface of the groove**. As shown in Fig. 2d, the EF intensity around the AuNR is enhanced by approximately 9 times compared to that on a bulk dielectric substrate (see Supplementary Note 3 and Supplementary Fig. S4 for details). When the SPR mode of the AuNR spectrally and spatially overlaps with the photonic mode, a strong spectrum localization is observed in the scattering response of the NR (Fig. 2e), characterized by a damping linewidth of 0.55 nm and a corresponding $Q$-factor of 1270, approximately 87 times higher than that located on a bulk dielectric (TiO$_2$) substrate. This indicates a substantial suppression of radiative losses in the SPR mode. In addition, our simulations show that the mode volume ($V_m$, see the Methods section and Supplementary Note 4 for details) of the AuNR SPRs on the FP-PCGR photonic substrate is reduced to approximately $7.96\times10^{-7}$ μm³, less than one-fifth of that on a dielectric substrate (approximately $4.33\times10^{-6}$ μm³) (Supplementary Figs. S5 and S6). This reduction in $V_m$ aligns well with recent simulations of hybrid plasmonic-photonic crystal systems[59], highlighting the approach's potential applicability to **a broader range of plasmonic nanostructures**. For example, substituting the AuNR with an Au nanodimer[60] further decreases $V_m$ to approximately $2.05\times10^{-7}$ μm³, while still maintaining a high SL with a quality factor $Q \sim 1270$ (see Supplementary Fig. S7). Additionally, **tuning the grating height** of the FP-PCGR photonic substrate provides a powerful means of controlling the damping linewidth of the substrate-engineered AuNR SPR mode. In particular, reducing the grating height allows the electric field surrounding the metal nanoparticle to extend further into the lossless air and low-loss photonic crystal. This shift effectively reduces metallic losses and leads to an enhancement in the spectral localization of the SPRs[25] (see Supplementary Fig. S8).

As evidenced in Fig. 2e, this phenomenon can be accurately predicted by our analytical calculations using Eq. (3), demonstrating that the profound spectral localization primarily arises from the distinct OP engineered by the photonic crystal substrate. This conclusion remains valid for the absorption spectrum of this substrate-engineered nanoparticle and the MNP positioned on the **top surface of the grating strip** in the FP-PCGR substrate (Supplementary Fig. S9), indicating the robustness of photonic substrate engineering in enabling strong SL at the single-nanoparticle level. Moreover, by varying the **thickness of the Au reflector** within the photonic crystal substrate, we can precisely control both the resonance linewidth and the spectral position of the OP, providing an additional handle to tailor the spectral localization of SPRs in individual MNPs (Supplementary Fig. S10). Fig. 2f further illustrates the tunability of the substrate-engineered SPR mode's wavelength by simply adjusting the **incident light angle**, demonstrating the versatility of this optoplasmonic system for applications in on-chip nanophotonic devices. This spectral localization strategy is general and can be experimentally realized through substrate engineering. To validate our approach, we performed **proof-of-concept experiments** for both "open" and "closed" optical pathways (OPs), implemented using two types of fabricated leaky Fabry–Pérot photonic substrates. These experiments collectively cover all four cases illustrated in Fig. 1c.

**Proof-of-concept experiment demonstrating "open" OP**

We first conduct a reference study of MNP in vacuum and on dielectric substrates, where no OPs exist in the electromagnetic environment (Supplementary Note 5 and Supplementary Fig. S11). In these cases, the SPRs are intrinsic plasmonic modes of the nanoparticles without OP reshaping. Then, we create an EM environment with OPs using a leaking FP cavity substrate made of Si/SiO$_2$/Si$_3$N$_4$, and integrate an AuNR on its top surface (Fig. 3a). In this photonic substrate, the thickness of the FP cavity is $t_{SiO2}$ = 3280 nm, corresponding to the thickness of the embedded SiO$_2$ layer. The flat surface of this substrate minimizes background scattering around the single nanoparticles, enabling precise collection of the scattering signals from the MNPs in dark-field scattering measurements (see Methods and Supplementary Fig. S12 for measurement setup). Driven by a weak plane wave, photonic modes leak into free space above the substrate, forming an EM environment around the AuNR. Fig. 3b shows the measured and simulated reflectance of the bare photonic substrate, revealing 5 photonic modes (TM$_1$-TM$_5$) in the spectral region of 550-750 nm. The red curve in Fig. 3b represents the calculated $F_m$ spectrum at $h$ = 12 nm above the substrate, exhibiting $F_m$ peaks (>1.0) with a Lorentzian line profile, indicating that these leaky photonic modes are bright and form open OPs. More variations in the $F_m$ spectrum at different heights can be found in Supplementary Fig. S13, illustrating the evolution of OPs engineered by this photonic substrate.

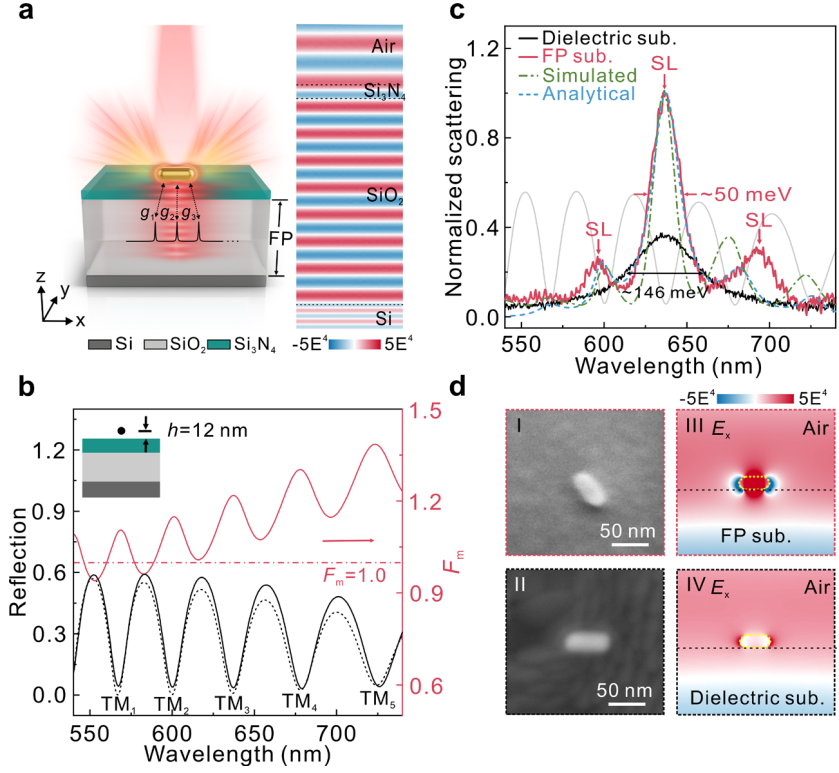

**Fig. 3 Proof-of-concept experiment of an AuNR on a photonic substrate with "open" OP.**
**a** Schematic of an AuNR on a leaking FP photonic substrate made of Si/SiO$_2$/Si$_3$N$_4$ (Left). Simulated EF distributions of this bare photonic substrate (Right). **b** Measured (solid black) and simulated (dashed black) reflectance of the photonic substrate with $t_{SiO2}$= 3280 nm and $t_{Si3N4}$ = 222 nm; the $F_m$ spectrum simulated at $h$ = 12 nm above the substrate surface (solid red). **c** Normalized measured scattering spectra from a single AuNR (52 nm in length and 12 nm in radius) on the FP photonic substrate (solid red) and the ITO-coated glass substrate (black), the simulated result (dashed olive), the analytical result (dashed blue) calculated using Eq. (3), and measured reflectance of the bare substrate (solid gray). **d** SEM images (I and II) of the measured AuNR, alongside the corresponding cross-sectional (*x-z*) view for EF distribution of the AuNR on the leaking FP (III) and dielectric (IV) substrates.

When individual AuNRs are placed on the top surface of this FP substrate (Supplementary Fig. S14), the synergistic interactions between the SPR mode, photonic modes, and vacuum reservoir significantly reshape their light scattering, as predicted by Eq. (3). Figure 3c showcases the normalized scattering measured from an individual AuNR, with its SPR mode resonant to the photonic mode at TM$_3$. The damping linewidth of this substrate-engineered AuNR is compressed to approximately 50 meV, down from about 146 meV for the AuNR located on the bulk ITO-coated glass substrate, and its intensity is enhanced by over 2 times, demonstrating clear spectral localization (**quadrant I**, Fig. 1c). In addition, the EF intensity on the AuNR's surface is increased by approximately 2 times compared to that on the bulk substrate (Fig. 3d). This experimental result aligns well with predictions from Eq. (3) and our simulations (Fig. 3c). Conversely, $TM_2$ and $TM_4$ represent the detuned photonic modes (**quadrant IV**, Fig. 1c), where the spectral localization with Lorentzian line shape can also be observed. By altering the **aspect ratio of the AuNR** or the **thickness of the SiO₂ layer** in the FP substrate, we can detune the SPR mode to different OPs, generating new plasmonic modes with varying intensities (Supplementary Fig. S15). This approach is universally applicable to various metallic nanostructures with **different geometries** (Supplementary Figs. S16 and S17).

**Proof-of-concept experiment demonstrating "closed" OP**

Inserting Au nanofilms as upper and lower reflectors in the leaking FP photonic substrate enables the switching of OPs from "open" to "closed" (Fig. 4a and Supplementary Fig. S18), thereby modifying the spectral shape of SPRs. Fig. 4b presents the reflectance and $F_m$ spectrum of this different FP photonic substrate composed of a $Si/Au/Si_3N_4/Au/Si_3N_4$ multilayer stack. The $F_m$ spectrum indicates that the OPs are "closed," resulting in $F_m < 1$ and Fano destruction for photonic modes $TM_1$-$TM_6$ (Fig. 4b and Supplementary Fig. S19). Although these photonic modes can couple with SPRs, the "closed" OPs cannot reshape the light scattering from the AuNR.

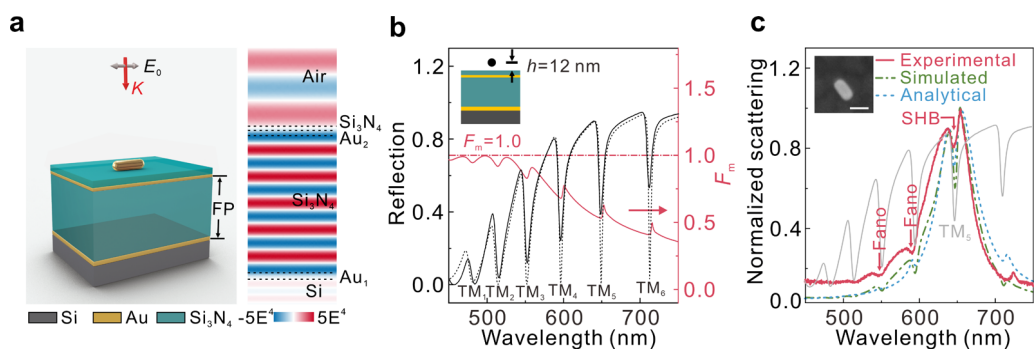

**Fig. 4 Proof-of-concept experiment of an AuNR on a photonic substrate with "closed" OP. a** Schematic of an AuNR on a FP photonic substrate made of $Si/Au/Si_3N_4/Au/Si_3N_4$ (Left).

Simulated EF distributions of this bare photonic substrate (Right). **b** Measured (solid black) and simulated (dashed black) reflectance of the photonic substrate made of Si/Au/Si$_3$N$_4$/Au/Si$_3$N$_4$, with Au-film thicknesses of 45 nm (upper) and 75 nm (lower). The cavity thickness within this substrate is $t_{Si3N4}$ = 1730 nm. The $F_m$ spectrum simulated at $h$ = 12 nm above the substrate surface (solid red). **c** Normalized scattering spectrum of a single AuNR measured on the photonic substrate (red), the simulated result (dashed olive), and the analytical result (dashed blue) calculated using Eq. (3). The solid gray line is the measured reflectance of the bare photonic substrate. Inset: SEM image of the sample (the scale bar is 50 nm).

Figure 4c shows the measured scattering spectrum of an AuNR on this substrate, confirming our mechanistic interpretation from Fig. 1c. Here, only the plasmonic mode is excited by the plane wave, which couples to the photonic mode that is disconnected from the vacuum reservoir. In contrast to Fig. 3d, the spectrum in Fig. 4c displays an SHB feature around TM$_5$ (the resonant case; **quadrant II** in Fig. 1c) and distinct Fano-resonance features around other photonic modes (the detuned cases; **quadrant III** in Fig. 1c), phenomena accurately predicted by Eq. (3) and supported by simulations. These effects are also observable in various Au nanostructures on the same substrate (Supplementary Figs. S20 and S21). While the SPR spectrum cannot be effectively localized, these behaviors provide valuable insights into Fano-resonance destruction across the optoplasmonic systems.

**Discussion**

To summarize, we have presented a comprehensive approach for achieving spectral localization of SPRs in single MNPs through photonic substrate engineering. This method leverages a mode-coupling theoretical framework that thoroughly considers the EM environment and engineered OPs. It elucidates the underlying mechanisms for manipulating the spectral profiles of single-nanoparticle SPRs across the four quadrants, manifesting as spectral localization, spectral hole burning, and Fano-resonance destruction. These phenomena are governed by the nature of the OPs—whether "open" or "closed"—and the resonance conditions between plasmonic and photonic modes. Our theory, simulations based on the FP-PCGR photonic crystal substrate, and proof-of-concept experiments using leaking FP photonic substrates collectively confirm that engineering OPs via photonic substrates is key to effectively controlling SPR behavior. This approach is modular, allowing for the flexible combination of various plasmonic nanoparticle designs with different substrate configurations. As a result, it serves as a foundational building block for advanced plasmonic nanodevices.

**Methods**

**Calculation of M-Factor**

We begin by applying the time-averaged form of Poynting's theorem to analyze the electromagnetic energy flux through an arbitrary surface. The theorem relates the time-averaged power outflow of EM energy, described by the time-averaged Poynting vector $\langle S \rangle = \frac{1}{2}\text{Re}(E \times H^*)$, to the fields and currents within the enclosed volume, as given by[54]

$$\langle P_{\text{out}} \rangle = \oiint \langle S \rangle \cdot n da = -\frac{1}{2}\int_V Re(j^* \cdot E)dV, \qquad (5)$$

where the current density is defined as $j = -i\omega p\delta(r - r_0)$, corresponding to a point dipole $p$ located at $r_0$. The electric field generated by this dipole can be expressed using the Green's function $G(r, r_0)$ of the system as: $E(r) = \omega^2 \mu G(r, r_0) \cdot p$. Substituting this into the expression for emitted power, we obtain:

$$P_{em} \propto \text{Im}(p^* \cdot E(r_0)) \propto p \cdot \text{Im}[G(r_0, r_0)] \cdot p \propto \rho(r, \omega, \hat{\mu}_d). \qquad (6)$$

We define the M-factor as the normalized projected local density of states (PLDOS): $F_m(r, \omega, \hat{\mu}_d) = \rho(r, \omega, \hat{\mu}_d)/\rho_0$. This quantity can also be directly computed using the ratio of the time-averaged power outflow in the system to that in vacuum[51]:

$$F_m(r, \omega, \hat{\mu}_d) = \frac{\langle P_{\text{out}} \rangle}{\langle P_{\text{out}} \rangle_0} = \frac{\oiint \langle S \rangle \cdot n da}{\oiint \langle S \rangle_0 \cdot n da}. \qquad (7)$$

Here, $\langle P_{\text{out}} \rangle = \oiint \langle S \rangle \cdot n da$ represents the time-averaged power outflow of electromagnetic energy through a closed surface[54] (see Supplementary Note 2 for details), and $\langle P_{\text{out}} \rangle_0$ denotes its corresponding value in vacuum.

**Calculation of mode volume $V_m$**

The mode volume $V_m$ of the SPR mode supported by Au nanostructures localized on photonic and dielectric substrates was calculated to quantify the spatial confinement of the EM energy, following the methods described in previous studies[62-64]. In dispersive media, the EM energy density is expressed as:

$$w(\mathbf{r}, \omega) = \frac{1}{2}\left[\frac{d[\omega\varepsilon_0\varepsilon(\mathbf{r},\omega)]}{d\omega}|\mathbf{E}(\mathbf{r},\omega)|^2 + \frac{d[\omega\mu_0\mu(\vec{r},\omega)]}{d\omega}|\mathbf{H}(\mathbf{r},\omega)|^2\right], \quad (8)$$

where $\varepsilon_0$ and $\mu_0$ are the vacuum permittivity and permeability, respectively, and $\varepsilon(\mathbf{r},\omega)$ and $\mu(\mathbf{r},\omega)$ are the relative permittivity and permeability at position $\mathbf{r}$. For lossy, dispersive materials such as metals, where the imaginary parts of permittivity and permeability are significant, the energy density expression is modified to:

$$w(\mathbf{r}, \omega) = \frac{\varepsilon_0}{2}\left[\varepsilon(\mathbf{r},\omega) + \frac{2\omega\mathrm{Im}[\varepsilon(\mathbf{r},\omega)]}{\gamma_p}\right]|\mathbf{E}(\mathbf{r},\omega)|^2 + \frac{\mu_0}{2}\left[\mu(\mathbf{r},\omega) + \frac{2\omega\mathrm{Im}[\mu(\mathbf{r},\omega)]}{\gamma_h}\right]|\mathbf{H}(\mathbf{r},\omega)|^2, \quad (9)$$

where $\gamma_p$ and $\gamma_h$ are the damping constants associated with the electric and magnetic responses, respectively. In the case of Au nanostructures, the energy contribution from the electric field dominates that of the magnetic field. Thus, the energy density can be approximated by the electric field contribution alone:

$$w(\mathbf{r}, \omega) \approx \frac{\varepsilon_0}{2}\left[\varepsilon(\mathbf{r},\omega) + \frac{2\omega\mathrm{Im}[\varepsilon(\mathbf{r},\omega)]}{\gamma_p}\right]|\mathbf{E}(\mathbf{r},\omega)|^2. \quad (10)$$

Finally, the mode volume of the SPR mode is calculated as:

$$V_m(\omega) = \frac{\int w(\mathbf{r},\omega)d^3r}{\max[w(\mathbf{r},\omega)]}. \quad (11)$$

This definition normalizes the total stored energy by the maximum local energy density, providing a physically meaningful measure of how confined the plasmonic mode is in space.

**Materials**

Gold(III) chloride trihydrate ($HAuCl_4 \cdot 3H_2O$, ≥99%), silver nitrate ($AgNO_3$, 99.8%), sodium borohydride ($NaBH_4$, 99%), cetyltrimethylammonium bromide (CTAB, ≥99%), benzylhexadecyldimethylammonium chloride (CTAC, 99%), L-ascorbic acid ($C_6H_8O_6$, >99%), potassium iodide (KI, 99%), sodium hydroxide (NaOH, 97%), trisodium citrate dihydrate ($C_6H_5O_7Na_3 \cdot 2H_2O$, 99%), and sodium chloride (NaCl, 99.5%) were purchased from Sigma-Aldrich. All solutions were prepared using deionized water with a resistivity of 18.25 MΩ·cm.

**Fabrication of Au nanoparticles and leaky Fabry–Pérot photonic substrates**

Gold nanoparticles, including nanorods (AuNRs) and nanospheres, were synthesized following the procedure reported in our previous work[61]. Bulk silicon substrates (MKNANO, Nanjing) were cleaned by sequential sonication in acetone, isopropyl alcohol, and deionized water for 10 minutes each. $SiO_2$ and $Si_3N_4$ layers were deposited via inductively coupled plasma chemical vapor deposition (ICPCVD) using the PlasmonPro System100 ICP180-CVD (Oxford Instruments). Au nanofilms used in the leaky Fabry–Pérot (FP) photonic substrates (Si/Au/$Si_3N_4$/Au/$Si_3N_4$) were deposited via electron beam evaporation (DE400 e-beam evaporator). Film thicknesses of both dielectric and Au layers in the photonic substrates (Si/$SiO_2$/$Si_3N_4$ and Si/Au/$Si_3N_4$/Au/$Si_3N_4$) were measured using a Filmetrics F20 film thickness measurement system (Filmetrics Inc., USA).

**Dark-field scattering measurements**

The schematic of the dark-field scattering measurement setup is shown in Supplementary Fig. S12. Prior to measurement, 5 μL of Au nanoparticle solution was drop-cast onto the surface of either a dielectric or photonic substrate. After 5 minutes, the droplet was removed, and the substrate was rinsed with deionized water and dried under nitrogen, isolating individual Au nanoparticles on the surface. To obtain the scattering spectra of single nanoparticles, dark-field optical microscopy was correlated with scanning electron microscopy (SEM) imaging of the same field of view. Scattering images and spectra were acquired using a dark-field microscope (Olympus BX53M, Olympus Inc., Japan) integrated with a monochromator (Acton Spectra Pro 2360, Acton Inc., USA), a 150 W quartz tungsten halogen lamp, and a cooled CCD camera (Princeton Instruments Pixis 100B_eXcelon, Acton Inc., USA) operating at −70 °C. To capture the scattering spectra of individual hybrids, light was introduced through a dark-field objective (100×, numerical aperture 0.80) and collected in the backward direction by the same objective. Color-scattering images were captured using a digital color camera (ARTCAM-300MI-C, ACH Technology Inc.) mounted at the imaging plane of the microscope.

**Acknowledgments**

This work was supported by the National Natural Science Foundation of China (Grant No. 12374326), the Key Project of the Natural Science Foundation of Henan Province (Grant No. 232300421141), the National Research Foundation, Singapore (Grant Nos. NRF-CRP26-2021-0004 and NRF-CRP31-0007), the Ministry of Education, Singapore (Grant No. MOE-T2EP50223-0001), the Agency for Science, Technology and Research, Singapore (Grant No. MTC IRG M24N7c0083), and the Singapore University of Technology and Design through the Kickstarter Initiative (Grant No. SKI 2021-04-12).

**Author Contributions**

R.M.L., W.L., and L.W. conceived the project and supervised the overall research. R.M.L., J.F.Z., W.J.Z., and J.F.L. developed the theoretical framework and carried out analytical calculations. J.F.Z., W.L., and G.X.L. conducted the FEM and FDTD simulations. S.H.F., W.L., and X.Y.F. were responsible for sample preparation. S.H.F., J.D.A., X.Y.F., and L.J.G. designed and performed the experimental setup and optical measurements. R.M.L., W.L., and L.W. led the interpretation of results and were primarily responsible for writing, revising, and finalizing the manuscript. All authors contributed to discussions, reviewed the data, and provided feedback on the manuscript.

**Competing Interests**

The authors declare that they have no competing interests.

**Additional Information**

Supplementary information is available in the online version of the paper. Correspondence and requests for materials should be addressed to RML, WL, or LW.

**Data Availability**

The data that support the plots within this paper and other findings of this study are available from the corresponding author upon reasonable request.

**Code Availability**

Codes used to generate simulated data are available from the corresponding author upon request.

**References**

1. Törmä, P. & Barnes, W. L. Strong coupling between surface plasmon polaritons and emitters: a review. *Rep. Prog. Phys.* **78**, 013901 (2015).
2. Gramotnev, D. K. & Bozhevolnyi, S. I. Plasmonics beyond the diffraction limit. *Nat. Photonics* **4**, 83–91 (2010).
3. Haffner, C. et al. Low-loss plasmon-assisted electro-optic modulator. *Nature* **556**, 483–486 (2018).
4. Guo, X. et al. Efficient all-optical plasmonic modulators with atomically thin Van Der Waals heterostructures. *Adv. Mater.* **32**, 1907105 (2020).
5. Brolo, A. G. Plasmonics for future biosensors. *Nat. Photonics* **6**, 709–713 (2012).
6. Rodrigo, O. et al. Mid-infrared plasmonic biosensing with graphene. *Science* **349**, 165–168 (2015).
7. M. Kauranen and A. V. Zayats, Nat. Photonics **6**, 737–748 (2012).
8. Nielsen, M. P., Shi, X., Dichtl, P., Maier, S. A. & Oulton, R. F. Giant nonlinear response at a plasmonic nanofocus drives efficient four-wave mixing. *Science* **358**, 1179-1181 (2017).
9. Fang, N., Lee, H., Sun, C. & Zhang, X. Sub-diffraction-limited optical imaging with a silver superlens. *Science* **308**, 534–537 (2005).
10. Willets, K. A., Wilson, A. J., Sundaresan, V. & Joshi, P. B. Super-resolution imaging and plasmonics. *Chem. Rev*. **117**, 7538–7582 (2017).
11. Talebian, S., Wallace, G. G., Schroeder, A., Stellacci, F. & Conde, J. Nanotechnology-based disinfectants and sensors for SARS-CoV-2. *Nature Nanotech*. **15**, 618–624 (2020).
12. Kumar, A., Kim, S., & Nam, J. –M. Plasmonically engineered nanoprobes for biomedical applications. *J. Am. Chem. Soc*. **138**, 14509–14525 (2016).
13. Khurgin, J. B. How to deal with the loss in plasmonics and metamaterials. *Nature*

*Nanotech*. **10**, 2–6 (2015).

14. Zhou, W. & Odom, T. W. Tunable subradiant lattice plasmons by out-of-plane dipolar interactions. *Nature Nanotech*. **6**, 423–427 (2011).

15. Boriskina, S. V. et al. Losses in plasmonics: from mitigating energy dissipation to embracing loss-enabled functionalities. *Adv. Opt. Photonics* **9**, 775-827 (2017).

16. Lu, H. -Y. et al. Extracting more light for vertical emission: high power continuous wave operation of 1.3-μm quantum-dot photonic-crystal surface-emitting laser based on a flat band. *Light Sci. Appl.* **8**, 108 (2019).

17. Kim, H., Noda, S., Song, B. -S. & Asano, T. Determination of nonlinear optical efficiencies of ultrahigh-Q photonic crystal nanocavities with structural imperfections. *ACS Photonics* **8**, 2839−2845 (2021).

18. Tang, H., Ni, X., Du, F., Srikrishna, V. & Mazur, E. On-chip light trapping in bilayer moiré photonic crystal slabs. *Appl. Phys. Lett*. **121**, 231702 (2022).

19. Oudich, M., Kong, X., Zhang, T., Qiu, C. & Jing, Y. Engineered moiré photonic and phononic superlattices. *Nat. Mater*. **23**, 1169–1178 (2024).

20. Wang, B. et al. Generating optical vortex beams by momentum-space polarization vortices centred at bound states in the continuum. *Nat. Photonics* **14**, 623–628 (2020).

21. Zhou, M. Increasing the Q-contrast in large photonic crystal slab resonators using bound-states-in-continuum. *ACS Photonics* **10**, 1519–1528 (2023).

22. Huang, C. et al. Ultrafast control of vortex microlasers. *Science* **367**, 1018–1021 (2020).

23. Liu, Z. et al. High-quasibound states in the continuum for nonlinear metasurfaces. *Phys. Rev. Lett*. **123**, 253901 (2019).

24. Meinzer, N., Barnes, W. L. & I, R. Hooper, Plasmonic meta-atoms and metasurfaces. *Nat. Photonics* **8**, 889–898 (2014).

25. Liang, Y., Tsai, D. P. & Kivshar, Y. From local to nonlocal high-plasmonic metasurfaces. *Phys. Rev. Lett*. **133**, 053801 (2024).

26. Bin-Alam, M. S. et al. Ultra-high-Q resonances in plasmonic metasurfaces. *Nat.*

*Commun*. **12**, 974 (2021).

27. Liang, Y. et al. Bound states in the continuum in anisotropic plasmonic metasurfaces. *Nano Lett*. **20**, 6351−6356 (2020).

28. Ahn, W., Hong, Y., Boriskina, S. V. & Reinhard. B. M. Demonstration of efficient on-chip photon transfer in self-assembled optoplasmonic networks. *ACS nano* **7**, 4470–4478 (2013).

29. Peng, P. et al. Enhancing coherent light-matter interactions through microcavity-engineered plasmonic resonances. *Phys. Rev. Lett*. **119**, 233901 (2017).

30. Gurlek, B., Sandoghdar, V. & Martín-Cano, D. Manipulation of quenching in nanoantenna–emitter systems enabled by external detuned cavities: a path to enhance strong-coupling. *ACS Photonics*, **5**, 456−461 (2018).

31. Doeleman, H. M., Verhagen, E. & Koenderink, A. F. Antenna–cavity hybrids: matching polar opposites for Purcell enhancements at any linewidth. *ACS Photonics* **3**, 1943−1951 (2016).

32. Wang, P. et al. Single-band 2-nm-line-width plasmon resonance in a strongly coupled Au nanorod. *Nano Lett*. **15**, 7581−7586 (2015).

33. Thakkar, N. et al. Sculpting fano resonances to control photonic–plasmonic hybridization. *Nano Lett*. **17**, 6927−6934 (2017).

34. Dezfouli, M. K., Gordon, R. & Hughes, S. Modal theory of modified spontaneous emission of a quantum emitter in a hybrid plasmonic photonic-crystal cavity system. *Phys. Rev. A* **95**, 013846 (2017).

35. Franke, S. et al. Quantization of quasinormal modes for open cavities and plasmonic cavity quantum electrodynamics. *Phys. Rev. Lett*. **122**, 213901 (2019).

36. Qian, Z. et al. Absorption reduction of large purcell enhancement enabled by topological state-led mode coupling. *Phys. Rev. Lett*. **126**, 023901 (2021).

37. Barth, M. et al. Nanoassembled plasmonic-photonic hybrid cavity for tailored light-matter coupling. *Nano Lett*. **10**, 891–895 (2010).

38. Doeleman, H. M., Dieleman, C. D., Mennes, C., Ehrler, B. & Koenderink, A. F. Observation of cooperative purcell enhancements in antenna–cavity hybrids. *ACS*

*Nano* **14**, 12027−12036 (2020).

39. Zhang, H. Y., Liu, Y. C., Wang, C. Y., Zhang, N. E. & Lu, C. C. Hybrid photonic-plasmonic nano-cavity with ultra-high Q/V. *Opt. Lett*. **45**. 4794 (2020).

40. Liu, J. –N., Huang, Q., Liu K. –K., Singamaneni, S. & Cunningham, B. T. Nanoantenna–Microcavity Hybrids with Highly Cooperative Plasmonic–Photonic Coupling. Nano Lett. **17**, 7569−7577 (2017).

41. Huang, Q. et al. Nanoantenna–Microcavity Hybrids with Highly Cooperative Plasmonic–Photonic Coupling. *ACS Photonics* **7**, 1994−2001 (2020).

42. Li, W. et al. Highly efficient single-exciton strong coupling with plasmons by lowering critical interaction strength at an exceptional point. *Phys. Rev. Lett*. **130**, 143601 (2023).

43. Zhou, N. et al. Strong mode coupling-enabled hybrid photon-plasmon laser with a microfiber-coupled nanorod. *Sci. Adv*. **8**, eabn2026 (2022).

44. Liu, J. et al. Reshaping plasmon modes by film interference. *Sci. China-Phys. Mech. Astron*. 66, 114211 (2023).

45. Hoang, T. B., Akselrod, G. M. & Mikkelsen, M. H. Ultrafast room-temperature single photon emission from quantum dots coupled to plasmonic nanocavities. *Nano Lett*. **16**, 270 (2016).

46. Andersen, S. K. H., Kumar, S. & Bozhevolnyi, S. I. Ultrabright linearly polarized photon generation from a nitrogen vacancy center in a nanocube dimer antenna. *Nano Lett*. **17**, 3889 (2017).

47. Mejía-Salazar, J. R. & Oliveira Jr, O. N. Plasmonic biosensing: Focus review. *Chem. Rev*. **118**, 10617−10625 (2018).

48. Yan, R. et al. Highly sensitive plasmonic nanorod hyperbolic metamaterial biosensor. *Photon. Res*. **10**, 84-95 (2022).

49. Liu, J. et al. A solid-state source of strongly entangled photon pairs with high brightness and indistinguishability. *Nature Nanotech*. **14**, 586–593 (2019).

50. Liu, R., Liao, Z., Yu, Y. -C. & X. -H. Wang, Relativity and diversity of strong coupling in coupled plasmon-exciton systems. *Phys. Rev. B* **103**, 235430 (2021).

51. Chen, G. Y. et al. Ab initio determination of local coupling interaction in arbitrary nanostructures: Application to photonic crystal slabs and cavities. *Phys. Rev. B* **87**, 195138 (2013).

52. Zubarev, D. N. Double-time Green functions in statistical physics. *Soviet. Phys. Usp*. **3**, 320–345 (1960).

53. Du, R. et al. How to obtain the correct Rabi splitting in a subwavelength interacting system. *Nano Lett*. **23**, 444–450 (2023).

54. Novotny, L. & Hecht, B. *Principles of Nano-Optics*. Cambridge University Press, 2012, ISBN: 9781139551793.

55. Moerner, W. E. & Bjorklund, G. C. Persistent spectral hole-burning: science and applications, Springer, Vol. 1 (1988).

56. Putz, S. et al. Spectral hole burning and its application in microwave photonics. *Nat. Photonics* **11**, 36–39 (2017).

57. You, J. -B. et al. Suppressing decoherence in quantum plasmonic systems by the spectral-hole-burning effect. *Phys. Rev. A* **103**, 053517 (2021).

58. Zhou, W. -J. et al. Cavity spectral-hole-burning to boost coherence in plasmon-emitter strong coupling systems. *Nanotechnology* **33**, 475001 (2022).

59. B. Fu, Enhanced light–matter interaction with Bloch surface wave modulated plasmonic nanocavities. *Nano Lett*. **25**, 722−729 (2025).

60. Liu, R. et al. Deterministic positioning and alignment of a single-molecule exciton in plasmonic nanodimer for strong coupling. *Nat. Commun*. **15**, 4103 (2024).

61. Liu, R. et al. On-demand shape and size purification of nanoparticle based on surface area. *Nanoscale* **6**, 13145-13153 (2014).

62. Chikkaraddy, R. et al. Single-molecule strong coupling at room temperature in plasmonic nanocavities. *Nature* **535**, 127–130 (2016).

63. G. Zengin, M. Wersäll, S. Nilsson, T. J. Antosiewicz, M. Käll, and T. Shegai, *Phys. Rev. Lett*. 114, 157401(2015).

64. R. Ruppin, Electromagnetic energy density in a dispersive and absorptive material**.** *Phys. Lett. A* **299**, 309-312 (2002).